\documentclass[prl,superscriptaddress, preprintnumbers,twocolumn,showpacs]{revtex4}
\usepackage{amsmath}
\usepackage{amssymb}
\usepackage{amsthm}
\usepackage{amsfonts}
\usepackage{algorithmic}
\usepackage{enumerate}
\usepackage{latexsym}
\usepackage{graphicx}
\usepackage{bm}
\usepackage{subfigure}
\usepackage{color}

\begin{document}

\title{Ferromagnetic barrier induced negative differential conductance on the surface of a topological insulator}

\author{Xing-Tao An}
\email{anxt@hku.hk}
\affiliation{School of Sciences, Hebei University of Science and Technology, Shijiazhuang, Hebei 050018, China}
\affiliation{Department of Physics and Center of Theoretical and Computational Physics, University of Hong Kong, Hong Kong, China}

\date{\today}

\begin{abstract}
We theoretically investigate the effect of the negative differential conductance of a ferromagnetic barrier on the surface of a topological insulator. Due to the changes of the shape and position of the Fermi surfaces in the ferromagnetic barrier, the transport processes can be divided into three kinds: the total, partial and blockade transmission mechanisms. The bias voltage can give rise to the transition of the transport processes from partial to blockade transmission mechanisms, which results in a giant effect of negative differential conductance. With appropriate structural parameters, the current-voltage characteristics show that the minimum value of the current can reach to zero in a wide range of the bias voltage, and a large peak-to-valley current ratio can be obtained.
\end{abstract}

\maketitle

In recent years, the new materials known as topological insulators have fueled tremendous research interest due to the rich fundamental physics and potential
applications. These materials have a fully insulating gap in the bulk and topologically protected gapless helical states on the boundary. These gapless states are completely immune to weak and nonmagnetic disorder and interactions due to the time-reversal symmetry and the topology of the bulk gap. Recent experiments have demonstrated the presence of such surface states in three-dimensional materials using angle-resolved photoemission spectroscopy,\cite{Hsieh} which have sparked significant interests on the subject from the point of view of both fundamental physics and potential benefits in numerous device applications. It has been found that the directions of the spin angular momentum and linear momentum are tied to each other in the surface states, in which electrons can be described by a massless Dirac equation. For practical applications, it is crucial to control the surface states as the time reversal symmetry is broken by using ferromagnetic materials. The exchange interaction with a proximate magnetic film can affect the surface states of topological insulators far more effectively than externally applied magnetic fields. Motivated by this understanding, several works have studied the effect of proximate ferromagnetic insulating film on the electron transport properties in topological insulators.\cite{Mondal, Wu, Yokoyama, Zhang, Kong, Zhai}

The effect of the negative differential conductance is well known to offer all possible applications including high-frequency oscillators, analog-to-digital converters, memory, and logic gates, etc.\cite{Mizuta} In semiconductor devices, such as semiconductor resonant tunneling diodes and superlattice structures, the negative differential conductance has been widely studied.\cite{Tsu, Sollner} In recent years, the negative differential conductance has also been explored and discussed in graphene nanostructures.\cite{Dragoman, Do1, Wang, Ren, Do2, Ferreira, Nguyen, Zhao, Song} The practical applications of the surface states of topological insulators, such as tunneling magnetoresistance,\cite{Yokoyama, Zhang, Kong} topological spin transistor\cite{Maciejko} and momentum filter\cite{Wu2} have been proposed in recent work. These lead to a further question: Can a negative differential conductance displays in a barrier on the surface of topological insulators?

In this Letter, we theoretically investigate the effect of the negative differential conductance induced by a ferromagnetic barrier on top of a topological surface. For a ferromagnetic barrier, a ferromagnetic insulator is put on the top of the topological insulator to induce an exchange field via the magnetic proximity effect. Meanwhile, a metal gate on the top of the insulator is used to control the electrostatic potential. We consider a realistic linear voltage drop between the source and drain electrodes, as shown in Fig.1 (a). The current-voltage characteristics can be calculated in a rotated spin space by using transfer matrix method within the Landauer-B$\ddot{u}$ttiker formalism. When the strength of exchange field is not larger than the electrostatic potential, the transmission of the ferromagnetic barrier evolves from partial to completely blocked transmission status as the bias voltage gradually increasing. Therefore, we can find that the negative differential conductance appears with appropriate structural parameters in this case. A large peak-to-valley current ratio can be be achieved in this ferromagnetic barrier on the surface of the topological insulator.

\begin{figure}[htb]
\includegraphics[scale=1.0,angle=0]{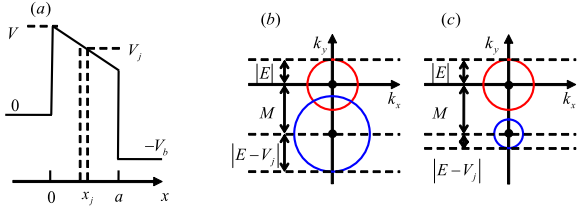}
\centering \caption{(color online) (a) Energy band diagram for a biased ferromagnetic barrier on the surface of a topological insulator. The overlap of the Fermi surfaces of the incident region and the region $x_{j}$ is plotted for (b) and  (c). The region $x_{j}$ is in the ferromagnetic barrier and its fermi surface (blue circle) has been shifted due to the induced exchange field $M$. $E$ is the energy of incident electron. $V_{j}$ is the potential in the $j$th layer and $a$ is the width of the ferromagnetic barrier.}
\label{figone}
\end{figure}

The motion of electrons on the topological insulator surface can be described by the two-dimensional Dirac Hamiltonian
\begin{eqnarray}
H=v_{f}{\bm p}\cdot{\bm \sigma}+{\bm M}\cdot{\bm \sigma}+V(x)\sigma_{0} \label{eq1},
\end{eqnarray}
where $v_{f}$ is the Fermi velocity, ${\bm p}=(p_{x} ,p_{y})$ is the electron momentum and $\sigma_{x}$ and $\sigma_{y}$ are Pauli matrices in spin space, ${\bm M}=M\hat{y}$ is the effective exchange field with magnetization aligned to the $y$ axis induced via magnetic proximity effect. $V(x)$ is the position-dependent electrostatic potential, and $\sigma_{0}$ is the $2\times2$ identity matrix. For convenience we express all quantities in dimensionless units by means of the length of the basic unit $l_{0}$ and the energy $E_{0}=\hbar{v}_{f}/l_{0}$. For a typical value of $l_{0}=50nm$ and the $Bi_{2}Se_{3}$ material $(v_{f}=5\times10^{5}m/s)$, one has $E_{0}=6.6meV$.

We describe the potential across the system as piecewise constant, as shown in Fig. 1(a). In each layer the potential is a constant $V_{j}=V-V_{bj}$, where $V$ is the electrostatic potential controlled by the top gate. The second term $V_{bj}=eV_{b}x_{j}/a$ is the potential energy drop across the $j$th layers due to the bias voltage $V_{b}$, where $x_{j}$ is the coordinate of the center of the layer $j$, and $a$ is the width of the ferromagnetic barrier. Because the system in our model is homogeneous along the $y$ direction, the transverse wavevector $k_{y}$ is conserved. The solution of the Dirac equation in the $j$th layer for a given incident energy $E$ can be expressed as
\begin{eqnarray}
\Psi_{j}&=&[c_{j}\left(
  \begin{array}{ccc}
    1\\
    S_{j}e^{i\theta_{j}}\\
  \end{array}
\right)e^{ik_{jx}x}\nonumber\\
&+&d_{j}\left(
  \begin{array}{ccc}
    1\\
    S_{j}e^{i(\pi-\theta_{j})}\\
  \end{array}
\right)e^{-ik_{jx}x}]e^{ik_{y}y}
 \label{eq2},
\end{eqnarray}
where $c_{j}$ and $d_{j}$ are the transmission and reflection coefficients, respectively. The positive (negative) sign of $S_{j}=sgn(E-V+V_{bj})$ corresponds to the quasiparticle in the electron-like (hole-like) region. $k_{jx}=\sqrt{(E-V+V_{bj})^{2}-(k_{y}+M)^{2}}$ is the longitudinal momentum in layer $j$ and $\theta_{j}=\tan^{-1}[(k_{y}+M)/k_{jx}]$.

Using the continuity of the wave function at the boundaries and the transfer matrix method, the transmission probability can be calculated $T=T(E,\theta)$, where $\theta$ is the incident angle of the electron. The ballistic conductance at zero temperature can be expressed in terms of the transmission probability
\begin{eqnarray}
G=G_{0}\int^{\pi/2}_{-\pi/2}T(E,\theta)\cos\theta{d}\theta \label{eq3},
\end{eqnarray}
where $G_{0}=e^{2}|E|L_{y}/(2\pi{h})$ is taken as the conductance unit, and $L_{y}\gg{a}$ is the length of the device in the $y$ direction. Then the net current
at zero temperature can be calculated
\begin{eqnarray}
I(V_{b})=I_{0}\int_{E_{f}-V_{b}}^{E_{f}}\int^{\pi/2}_{-\pi/2}T(E,\theta,V_{b})|E|\cos\theta{d}\theta{d}E \label{eq4},
\end{eqnarray}
where $I_{0}=ev_{f}L_{y}/(2\pi{l_{0}})^{2}$ is a current unit and $E_{f}$ is the Fermi energy of the system.

The Fermi surfaces in the incident region (red circle) and in the ferromagnetic barrier region (blue circle) are plotted in Fig. 1(b) and (c). The sizes of the Fermi surfaces in the ferromagnetic barrier region have been changed due to the position-dependent electrostatic potential, and the positions of those in the momentum space move to the $k_{y}$ direction due to the exchange field. This misfit of the in-plane momentum between the incident region and the ferromagnetic barrier region gives rise to a strong dependence of the current on the bias voltage. Since $k_{y}$ is conserved, the permitted incident angle of the electron is determined by the overlap of the $k_{y}$ ranges of the two regions. When $|V(x)-E|-M>|E|$ or $|E|-M>|V(x)-E|$, all the incident angles are allowed and the ferromagnetic barrier is in the total transmission status. When $|V(x)-E|-M<|E|$ and $|V(x)-E|+|E|>M$, the overlap between the Fermi surfaces in the two regions is reduced and hence a fraction of incident angle is allowed to transmit, as shown in Fig 1(b). In this partial transmission status, the conductance is strongly suppressed. When $|V(x)-E|+|E|<M$, the overlap between the $k_{y}$ ranges of the two regions will be absent, as shown in Fig. 1(c). This is the blockade status, in which the conductance is completely suppressed.

In the following, we will show results for the electrostatic potential $V=5$. The conductance strongly depends on the bias voltage, which will be explained in Fig. 2 for various bias voltage. In Fig. 2, we show the conductance $G$ of the system versus the incident energy $E$. To describe the physics, we first consider the $V_{b}=0$ case (solid curves in Fig. 2). We can see that the conductance decreases when the incident energy is less than the electrostatic potential ($E<V$) and then increases when the incident energy increases. When the incident energy equals the electrostatic potential, the conductance reaches the minimum value due to $k_{x}^{2}+(k_{y}+M)^{2}=(E-V)^{2}$. This effect and the oscillatory behavior of the conductance become stronger when the width of the ferromagnetic barrier increases. In the case of $V>M$, the system is in the total transmission status as $E<(V-M)/2$ and $E>(V+M)/2$, and is in the partial transmission status as $(V-M)/2<E<(V+M)/2$. With the bias voltage applied, the potential in layer $j$ becomes position-dependent potential $V-V_{bj}$. It is interesting to note that the total transmission is absent when $V-V_{b}<M$, in which the system is in the partial transmission status as $E<(V-V_{b}-M)/2$ and $E>(V-V_{b}+M)/2$, and is in the blockade status as $(V-V_{b}-M)/2<E<(V-V_{b}+M)/2$. Therefore, as shown in Fig. 2, the system will evolve from the partial transmission status to the blockade status as the bias voltage gradually increasing. This evolvement can give rise to the effect of the negative differential conductance, which is confirmed by our following discussion.

\begin{figure}[htb]
\centering
\includegraphics[scale=1.0,angle=0]{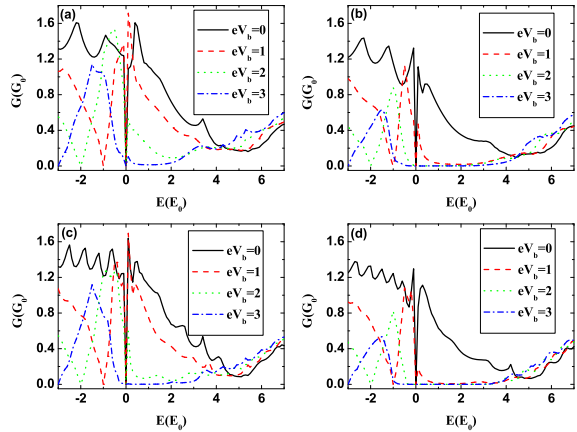}
\caption{(Color online) The conductance as a function of the energy $E$ with various bias $V_{b}$ for (a) $a=2$ and $M=3$, (b) $a=2$ and $M=4$, (c) $a=4$ and $M=3$, (d) $a=4$ and $M=4$.} \label{figtwo}
\end{figure}

In order to further investigate this effect of the bias voltage on the transport properties of the system, we calculate the transmission probability $T$ as a function of the incident angle of the electron for $E=2$, as shown in Fig. 3. In the case of $V_{b}=0$, the transmission probability exhibits resonant tunneling behavior which becomes stronger with the width of the ferromagnetic barrier increasing. The incident direction of Klein tunneling is not normal incidence but is changed by the exchange field. When $M=3$ [Fig. 3(a) and (c)], we can see that a half of electrons with the incident angle $-\pi/2<\theta<0$ are allowed to transmit as $V_{b}=0$. This is because only the negative $k_{y}$ is in the overlap of the Fermi surfaces of the incident region and the ferromagnetic barrier region as $M=|E-V|$, which can be made clear in Fig. 1(b). Increasing the exchange field $M$, for example $M=4$ in Fig. 3(b) and (d), the region of resonant tunneling decreases because of the decreasing of the overlap of the two Fermi surfaces. For the case of $V_{b}\neq0$, the bias voltage can reduce the value of $|E-V_{j}|$ and then reduce the overlap between the $k_{y}$ ranges of the incident and the ferromagnetic barrier regions when the incident energy $E$ and the exchange field $M$ are fixed, which can be seen in Fig. 1(b) and (c). Therefore, the transmission probability is suppressed by the bias voltage, as shown in Fig. 3. Especially, in Fig. 3(d), we can see that all of the incident electrons are not allowed to transmit and the system is in the blockade status for $eV_{b}=2$ and $eV_{b}=3$.

\begin{figure}[htb]
\centering
\includegraphics[scale=1.0,angle=0]{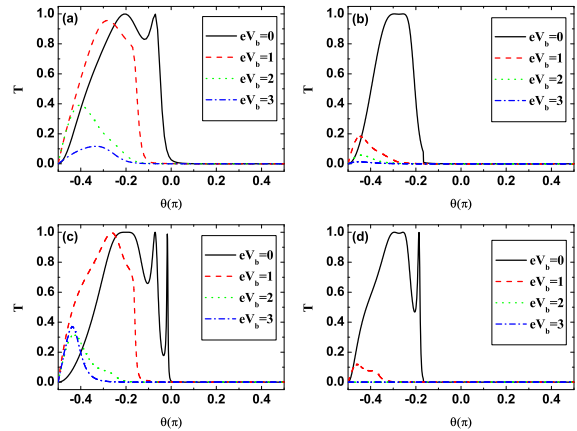}
\caption{(Color online) The transmission probability $T$ as a function of the incident angle with various bias $V_{b}$ for $E=2$. Other parameters are the same as those of Fig. 2.} \label{figthree}
\end{figure}

Finally, we show the current-voltage characteristics of the system at various Fermi energies in Fig. 4. It is clear that obvious negative differential conductance behaviors are observed. The currents in the system increase linearly under low bias and then drop dramatically. The form of the current-voltage curve can easily be explained by looking at the conductance curve with various $V_{b}$ in Figs. 2 and the overlap of the Fermi surfaces in Fig. 1: at low bias, the contribution of the partial transmission processes raises the current; when further raising the bias, the partial transmission is suppressed and the bias voltage bring the system into the blockade status, and therefore the current decreases; when the bias is high enough, the contribution of the electrons in valence band causes a very rapidly rising current because the integration window in Eq. (4) is $[E_{f}-eV_{b}, E_{f}]$ rather than $[Max(E_{c}, E_{f}-eV_{b}), E_{f}]$ in normal semiconductors. Here, $E_{c}$ is the conduction band bottom of normal semiconductors and $Max(u,v)$ stands for the bigger one of $u$ and $v$. It is shown that the negative differential conductance behavior appears more clearly and its peak-to-valley current ratio increases when the exchange field $M$ and the width of the ferromagnetic barrier $a$ increase, although the current peak is reduced. Very interestingly, the current is quenched to zero in the range of $1.5E_{0}/e<V_{b}<4E_{0}/e$ in the case of $E_{f}=3$ for $M=4$ and $a=4$, as shown in Fig. 4(d). As a result, a large peak-to-valley current ratio can be obtained. This giant effect of negative differential conductance in the ferromagnetic barrier on the surface of topological insulators is exactly the aim of the present work.

\begin{figure}[htb]
\centering
\includegraphics[scale=1.0,angle=0]{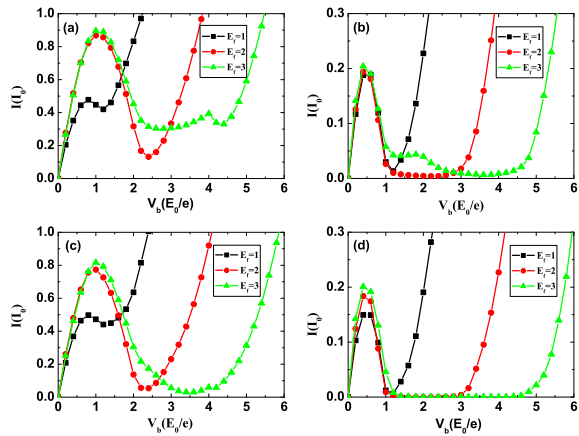}
\caption{(Color online) Current-voltage characteristics of the ferromagnetic barrier on the surface of topological insulators for the different Fermi energies $E_{f}$. Other parameters are the same as those of Fig. 2.} \label{figfour}
\end{figure}

In summary, we study the effect of the negative differential conductance induced by a ferromagnetic barrier on the surface of topological insulators. We find
that the transmission processes of the system can be classified three types: the total, partial and blockade transmission status according to the competition between the electrostatic potential and the exchange field. With the bias voltage gradually applied, the system will evolve from the partial transmission status to the blockade status, which can give rise to the effect of the negative differential conductance. The current-voltage characteristics show that the minimum value of the current can reach to zero in a wide range of the bias voltage with appropriate structural parameters, and a large peak-to-valley current ratio can be obtained. The origin of this negative differential conductance is different from either that in the normal semiconductors or that in the gaped graphene.
This negative differential conductance completely arises from the transition of the transmission processes induced by the bias voltage. The results obtained in the present study may have certain practical significance in applications for future topological insulator electronic devices.

This work was supported by National Natural Science Foundation of China (Grant Nos. 11104059 and 61176089) and Hebei province Natural Science Foundation of China (Grant No. A2011208010).

\end{document}